\documentstyle[11pt,twoside,sympo2011,epsfig]{article}
\def\corot {CoRoT}

\def\cesam {{\sc{cesam}}}
\def\graco {{\sc{graco}}}

\def\dsct {$\delta$ Sct}

\def\starb {HD\,174966}
\def\strom {Str\"{o}mgren}
\def\feh {[Fe/H]}
\def\logg {$\log g$}
\def\mv {$m_{\mathrm{v}}$}

\def\vsini {$v\cdot\!\sin\!~i$}
\def\kms {$\mathrm{km}\,\mathrm{s}^{-1}$}
\def\teff {$T_{\mathrm{eff}}$}

\def\muhz {$\mu\mbox{Hz}$}

\def\Dnu {$\Delta\nu$}

\begin{document}
\title{Quasiperiodic patterns in $\delta$ Scuti stars: an in depth study of the CoRoT star HD 174966}
\author{A. Garc\'\i a Hern\'andez$^1$, A. Moya$^2$, J. C. Su\'arez$^1$, R. Garrido$^1$, L. Mantegazza$^3$, S. Mart\'\i n-Ruiz$^1$, M. Rainer$^3$, E. Poretti$^3$, P. J. Amado$^1$, A. Rolland$^1$, P. Mathias$^4$, K. Uytterhoeven$^5$}
\affil{$^1$ Instituto de Astrof\'\i sica de Andaluc\'\i a (CSIC), Glorieta de la Astronom\'\i a s/n, 18008, Granada, Spain [agh@iaa.es]\\$^2$ Laboratorio de Astrof\'\i sica Estelar y Exoplanetas, LAEX-CAB (INTA-CSIC), PO BOX 78, 28691 Villanueva de la Ca\~nada, Madrid, Spain\\$^3$ INAF - Osservatorio Astronomico di Brera, via E. Bianchi 46, 23807 Merate (LC), Italy\\$^4$ Lab. d'Astrophysique de Toulouse-Tarbes, Universit\'e de Toulouse, CNRS, 57 Avenue d'Azereix, 65000 Tarbes, France\\$^5$ Instituto de Astrofísica de Canarias, 38200 La Laguna, Tenerife, Spain} 
\begin{abstract}
In this work, we have gone one step further from the study presented in the first \corot\ symposium. 
Our analysis consists on constructing a model database covering the entire uncertainty box of the \dsct\ star \starb, derived from the usual observables (\teff, \logg\ and \feh), and constraining the models representative of the star. To do that, we use the value of the periodicity (related to $\Delta\nu_{\ell}$) found in its \corot\ pulsating spectrum.
\end{abstract}
\section{Introduction and data description}
\starb\ is a \dsct\ A3 type main-sequence star of \mv=7.72. The physical parameters of this object are taken from the {\it{\corot\ Sky Database}} (Charpinet et al., 2006): \teff\ = 7637 $\pm$ 200 K, \logg\ = 4.03 $\pm$ 0.2 dex, \feh\ = -0.11 $\pm$ 0.2 dex and \vsini\ = 125 \kms. Our starting point is the method developed in our previous work (Garc\'\i a Hern\'andez et al. 2009, from now AGH2009): we are going to use the information so obtained to reduce the uncertainty in the determination of the physical parameters of this star.\par
We have used in this work the \corot\ light curve, new \strom\ photometry observations and spectroscopic data. \starb\ was observed during the first 30-day {\it{short run}} (SRc01) of \corot, obtaining a total of 73440 points. After data reduction, we have extracted a total of 185 independent frequencies, taken a conservative criteria for the spectral significance ({\it{sig}}) of 10.\par
The \strom\ photometry observations were carried out along 29 nights with the 90 cm telescope at the Observatorio de Sierra Nevada (OSN). A total of 3 frequencies were extracted from the light curves, which correspond with the 3 highest frequencies obtained from the \corot\ data. The spectroscopic observations were carried out along a total of 53 days using three different facilities: HARPS, at the La Silla observatory (ESO); FOCES, at the Calar Alto Observatory; and SOPHIE, the spectrograph of the Haute Provence Observatory. A Line Profile Variation (LPV) analysis was developed in order to obtain a modal identification. A total of 17 frequencies were detected, from which 4 correspond to the highest ones extracted from the \corot\ data.\par

\section{Results and discussion}

Following the method described in AGH2009, we have looked for a quasiperiodic pattern in the frequency set of \starb. We started by selecting different subsets containing the highest frequencies: 30, 60, 150 and, finally, all the spectrum. Then, we calculated a Fourier transform (FT) of the position of these frequencies, not taken into account the information of the amplitude (we took a standard amplitude equal 1). We found in these FT's the clear sign of a periodicity with a value of around 64~\muhz. We shown in AGH2009 that this periodicity can be seen in the models and that it correspond to a flattened zone in the growth of the large separation value (defined as usual $\Delta\nu_{\ell}=\nu_{n+1,\ell}-\nu_{n,\ell}$), which occurs around the fundamental radial mode.\par
Nevertheless, this type of pulsators usually has high rotational velocities, so the effects of such velocities in the oscillation modes are not negligible. At this values in the velocities, the perturbative theory is not enough to understand the whole frequency spectrum observed. Ligni\`eres et al. (2010) used a non-perturbative theory to calculate the pulsating frequencies of a fast rotating polytropic model. They found that the autocorrelation function of the frequency set shows, under certain conditions of the inclination angle of the star and visibility of the modes, a peak corresponding to the double of the rotational velocity.\par
For the case of \starb, an estimation of the inclination angle is derived from the spectroscopic observations: $i$ = 62.5$\deg$. Constructing the uncertainty box for this object, we can calculate the minimum and maximum radius and the limits in the possible rotational splitting: $\Delta_{split}=[15,31]$~\muhz. So, we cannot discard, at a first glance, the rotational origin of the periodicity found in the frequency spectrum of this pulsator. However, the models used in the work by Ligni\`eres et al. have rotational velocities of around a 60\% of the critical Keplerian velocity ($\Omega_{K}$). Using the measured values, \starb\ had not more than a 35\% of  $\Omega_{K}$. This ``slow'' rotation rate velocity makes more favorable to see \Dnu\ in the autocorrelation function (and in a FT of the frequency spectrum). Even more, the large separation value does not change too much from a non-rotating model to a high rotating one (Ligni\`eres, 2006).\par
We can then use non-rotating models and the large separation value extracted from the frequency set in order to reduce the uncertainty in the determination of the physical parameters of \starb. To do that, we have computed a non-rotating equilibrium and pulsating model database covering the $1\sigma$ uncertainty box, for \teff, \logg\ and \feh. The database was computed using the codes \cesam\ (Morel, 1997), for the equilibrium models, and \graco\ (Moya et al., 2004), for the pulsating ones. Using the value of the large separation derived from the \corot\ observations with a conservative uncertainty in its determination, we reduce the number of the representative models of the star. A value of \Dnu$_{\ell}$ = [60,80]~\muhz\ is taken, which means a reduction of a 76\% in the number of models. This translates too in a reduction of the uncertainties in, for example, the mass, the mean density and the age of 51\%, 56\% and 18\%, respectively.\par
We could go further in our analysis and use the information on the modal identification, both photometric and spectroscopic. Doing that, the number of models constraining all the measures are only 8. But the modal identification is strongly influenced by the rotation of the star, which we have not taken into account. So this last result should be taken with care.\par

\section{Conclusions}

We have carried out an in depth analysis of the \dsct\ star \starb. We have obtained 185 frequencies from the \corot\ light curve and, using a FT, we have found a periodic pattern corresponding to a large separation structure of \Dnu = [60,80]~\muhz. We have used the \Dnu\ information to reduce the uncertainties in the physical parameters of this pulsator. For example, we have reached a reduction of a 51\% for the mass, 56\% for the mean density and 18\% for the age.

%
%
%
%
%
%
%

\end{document}